\documentclass[aps,pre,preprint,groupedaddress,showpacs]{revtex4-1}

\usepackage{graphicx}
\usepackage{amsfonts}

\newcommand{\dd}{{\rm d}}

\begin{document}

\title{Non-equilibrium Entropy and Dynamics in a System with Long-Range Interactions}
\author{T.~M.~Rocha Filho}
\affiliation{Instituto de F\'\i{}sica and International Center for Condensed Matter Physics\\ Universidade de
Bras\'\i{}lia, CP: 04455, 70919-970 - Bras\'\i{}lia, Brazil}
\begin{abstract}
We extend the core-halo approach of Levin et al.\ [Phys.\ Rep.\ {\bf 535}, 1 (2014)] for the violent  relaxation of long-range interacting system
with a waterbag initial conditions, in the case of a widely studied Hamiltonian Mean Field
model. The Gibbs entropy maximization principle is considered with the constraints of energy conservation and of coarse-grained
Casimir invariants of the Vlasov equation. The core-halo distribution function depends only on the one-particle mean-field energy, as is expected
from Jeans Theorem, and depends on a set of parameters which in our approach are completely determined without having to solve
an envelope equation for the contour of the initial state, as required in the original approach. We also show that a different ansatz can be
used for the core-halo distribution with similar results.
This work also evidences a link between a parametric resonance causing the non-equilibrium phase transition in the model,
a dynamical property, and a discontinuity of the (non-equilibrium) entropy of the system.
\end{abstract}

\maketitle

\section{Introduction}

If the  pair-interaction potential of a many-body system
$V({\bf r}-{\bf r}^\prime)$ decays at long distances $r=|{\bf r}-{\bf r}^\prime|$ as
$1/r^\alpha$ with $\alpha$ smaller than the spatial dimension, then the system is said to be long-range interacting.
This includes the relevant cases of gravitational and non-shielded Coulomb interactions. These systems
present some unusual behavior when compared to short-range interacting systems, such as negative specific heat in the
microcanonical ensemble, non-ergodicity and non-Gaussian Quasi-Stationary States (QSS)~\cite{newbook}. Starting from an initial configuration, an isolated
many-particle system with long-range interactions evolves rapidly though a violent relaxation into a QSS 
which relaxes to equilibrium with a characteristic time diverging with the number of particles $N$~\cite{physrep,dynamics}, or in some cases oscillates around a QSS~\cite{konishi}.
Predicting the outcome of the violent relaxation has been a major problem in astrophysics for at least half a century, and has drawn much attention
in plasma physics and related fields. The first attempt for a statistical theory of violent relaxation is due to Lynden-Bell~\cite{lyndenbell} and is based on the Vlasov
equation description,
valid for short times for the one-particle distribution function~\cite{braun,dynamics}. For a system of identical particles with mass $m$ it is given by:
\begin{equation}
\dot f=\frac{\partial f}{\partial t}+\frac{\bf p}{m}\cdot\frac{\partial f}{\partial\bf r}+{\bf F}({\bf r},t)\cdot\frac{\partial f}{\partial\bf p}=0,
\label{vlasoveq}
\end{equation}
where ${\bf r}$ and ${\bf p}$ are the position and momentum vectors respectively, $f\equiv f({\bf r},{\bf p},t)$ the one-particle
distribution function, and the mean-field force at position ${\bf r}$ and time $t$ is:
\begin{equation}
{\bf F}({\bf r},t)=-\nabla\overline{V}({\bf r},t),
\hspace{5mm}\overline{V}({\bf r},t)\equiv\int V\left({\bf r}-{\bf r}^\prime\right)f({\bf r}^\prime,{\bf p}^\prime,t)\:d{\bf r}^\prime d{\bf p}^\prime,
\label{mfforce}
\end{equation}
where $\overline{V}({\bf r},t)$ is the mean-field potential.
The Vlasov equation admits infinitely many Casimir invariants of motion of the form
\begin{equation}
C_s[f]=\int s(f({\bf r},{\bf p},t))\:d{\bf r}d{\bf p},
\label{casimirs}
\end{equation}
for any function $s$. Setting $s=-k_B f\ln f$ in Eq.~(\ref{casimirs}) results in the Boltzmann entropy (with $k_B$ the Boltzmann constant) which is constant as
the Vlasov equation is reversible.
Assuming a complete mixing of micro-cells of same $f$-levels into coarse grained macro-cells, and maximizing the entropy given by the logarithm of
the number of possibilities of distributing non-overlapping microcells ($f$ is constant along phase trajectories) into all macro-cells, Lynden-Bell determined
the distribution function resulting from a violent relaxation. Although elegant, this approach is only valid as a first
approximation, as pointed out by Lynden-Bell himself (see~\cite{bindoni} for a review in astrophysics applications). Different tentative improvements were proposed in
the literature by Shu~\cite{shu}, Kull, Treumann and B\"ohringer~\cite{kull} and Nakamura~\cite{nakamura}, although none proved to be completely satisfactory~\cite{arad1,arad2}.

Parametric resonance is known to play a major role in plasmas as an important phenomenon in the long-time evolution of the system, and has been studied in the context of
a single wave-particle interaction~\cite{firpo1,firpo2}. Taking into account this mechanism, and
using Jeans theorem which states that in a steady state the distribution function $f$ depends on position
and momentum only through constants of motion~\cite{binney}, Levin and collaborators~\cite{levin1,levin2,levin3,levin4,levin5,levin6,levin7,levipprep}
proposed that for a waterbag initial distribution of the form
\begin{equation}
f({\bf r},{\bf p},0)=\eta\,\Theta(r_0-|{\bf r}|)\Theta(p_0-|{\bf p}|),
\label{wbdist}
\end{equation}
with $\eta$ a normalization constant and $\Theta$ the Heaviside step function, the final distribution function after
violent relaxation assumes a core-halo structure given by:
\begin{equation}
f({\bf r},{\bf p},t_v)=\eta\,\Theta(e_F-e({\bf r},{\bf p}))+\chi\,\Theta(e({\bf r},{\bf p})-e_F)\Theta(e_H-e({\bf r},{\bf p})),
\label{corehalodist}
\end{equation}
with $e({\bf r},{\bf p})$ the one-particle energy:
\begin{equation}
e({\bf r},{\bf p})=\frac{p^2}{2m}+\overline{V}({\bf r}),
\label{mfenergy}
\end{equation}
where $e_F$ and $e_H$ are called the Fermi and halo energies, respectively, $\chi$ is
a constant parameter fixed by the normalization condition and $\eta$ has the same value as the initial condition $f$-value in Eq.~(\ref{wbdist}).
A similar type of distribution function was used in Ref.~\cite{antoniazzi0} in the study of the free electron laser, by applying the approach of
Ref.~\cite{julien} which derives from the original Lynden-Bell theory for violent relaxation.

The characteristic time for violent relaxation $t_v$ is defined as the mean-field relaxation time into a stationary state of the Vlasov dynamics,
which is much shorter than the collisional relaxation to thermodynamic equilibrium (see~\cite{lyndenbell} for an estimation of $t_v$ for
a self-gravitating system).
The first term in the right-hand side of Eq.~(\ref{corehalodist}) corresponds to the core of the distribution and the second term
to the halo. The halo is populated by particles expelled by a parametric resonance in wave-particle interactions which injects particles at higher energies. The
halo energy $e_H$ in Eq.~(\ref{corehalodist}) corresponds to the highest energy attained by resonant particles, which in the original approach is determined by solving an
envelope equation for the evolution of the contour of the initial waterbag distribution~\cite{levipprep}.
This approach was applied with reasonable success for one- and two-dimensional
self-gravitating systems, non-neutral plasmas and the Hamiltonian Mean Field (HMF) model (see~\cite{levin1} and references therein).

Despite its success, the core-halo approach still requires to determine $e_H$ by solving a dynamical equation (the envelope equation),
which is not always a simple task, or by determining the halo energy from a Molecular Dynamics (MD) simulation,
in contrast to Lynden-Bell's approach that only presupposes a good mixing of phase elements (values of the distribution function $f$).
One of the goals of this paper is to show that by using an entropy maximization principle, no dynamical equation(s) have to be explicitly solved for.
We illustrate our approach by applying it to the HMF model that has played an important role as a paradigmatic model,
solvable at equilibrium and
retaining some important features of the dynamics of long-range interacting systems, yet being simple enough to allow large-scale
MD simulations with numeric effort scaling with the number of particles $N$ instead of $N^2$ as
for most systems of interest~\cite{nv1,nv2,nv3,jain,buyl,hmforig,physrep}.

The structure of the paper is as follows:
In the next section we present the HMF model and study its non-equilibrium phase-diagram for an initial waterbag state
by applying to it the Core-Halo approach as described by Pakter and Levin~\cite{levin4}.
In Section~\ref{varmethod} we introduce the variational approach as an alternative method
to determine the remaining parameter $e_H$ in the core-halo distribution as given in Eq.~(\ref{corehalodist}), and also discuss
the possibility of using a different ansatz for the core-halo configuration.
We close the paper with some concluding remarks in Section~\ref{conclusions}.

\section{The Hamiltonian Mean Field Model}
\label{hmfmodel}

\begin{figure}[ptb]
\begin{center}
\scalebox{1.0}{{\includegraphics{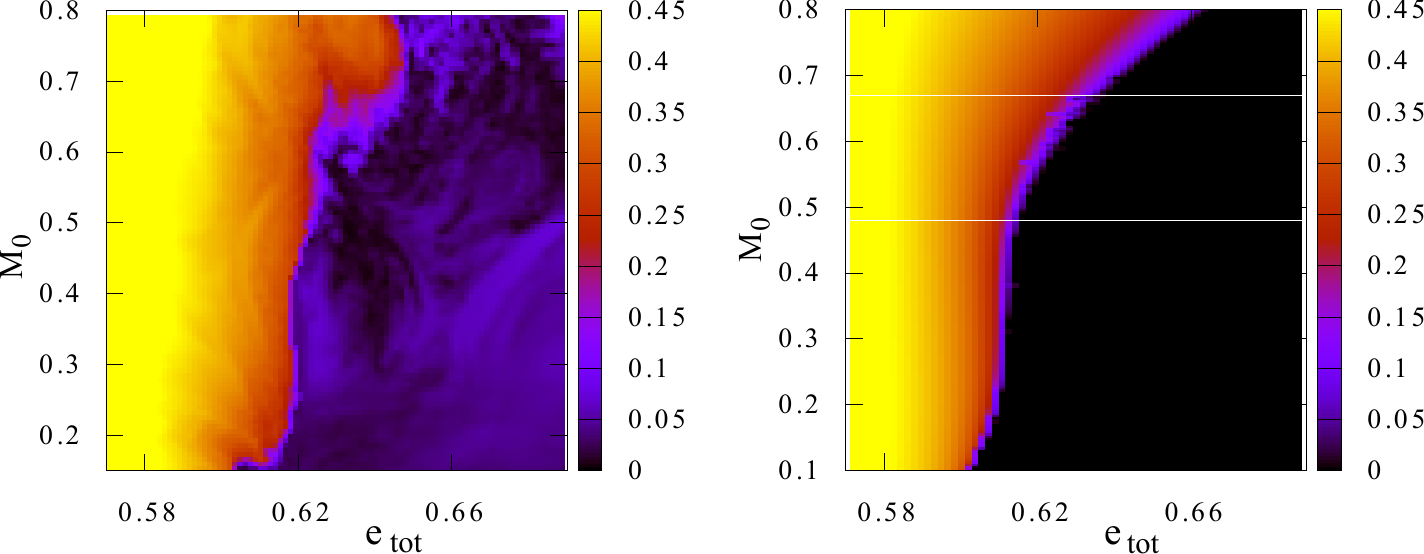}}}
\end{center}
\caption{(Color online) Left panel: Final magnetizations for the HMF model as a function of initial
magnetization $M_0$ and total energy per particle $e_{\rm tot}$ from numeric solutions of the Vlasov equation.
The grid in $(M_0,e_{\rm tot})$ space is formed by $100\times100$ simulations with $t_f=1000.0$ and a numerical grid of $1024\times1024$ points.
The final magnetization was obtained by averaging from $t=800.0$ to $t=1000.0$. Right panel: predictions from Lynden-Bell theory.}
\label{fig1}
\end{figure}

The Hamiltonian for the  HMF model is given by~\cite{hmforig}:
\begin{equation}
H=\sum_{i=1}^N \frac{p_i^2}{2}+\frac{1}{N}\sum_{i<j=1}^N \left[1-\cos(\theta_i-\theta_j)\right],
\label{hmfham}
\end{equation}
with $\theta_i$ the position angle of particle $i$ and $p_i$ its conjugate momentum.
At equilibrium, it exhibits a phase transition from a paramagnetic (homogeneous) phase at higher energies
to a ferromagnetic (non-homogeneous) phase at lower energies. For non-equilibrium states,
a similar phase transition is also observed but with a more intricate structure with a first or second order transition
and phase reentrances depending on the initial energy and magnetization, which do not coincide
with predictions from Lynden-Bell theory~\cite{maa7a,maa7b,maa7d,maa7e}.
Using large-scale molecular dynamics simulations and numerical solutions of the Vlasov equation~\cite{eu2,eu1},
the author and collaborators have shown that this phase structure is in fact much more complex than previously described,
with cascades of phase reentrances near the discontinuous phase-transition~\cite{nosnoneq}.
Non-equilibrium phase transitions in this model can be studied by solving the Hamiltonian equations of motion by using
a numerical integrator or solving the corresponding Vlasov equation. For the latter case, a semi-Lagrangian second order method in time
with a fixed time step is used (see Ref.~\cite{eu2} for details),
and integrated for a sufficiently long time such that after the initial waterbag state the system
has settled down in a QSS up to some long-lasting oscillations~\cite{nosnoneq}.
Figure~\ref{fig1} shows in greater detail than previous results the non-equilibrium phase diagram of the HMF
model from the numeric solution of the Vlasov equation and from Lynden-Bell theory~\cite{antoniazzi}.
Although predicting within some accuracy the transition line in the $(M,e)$ plane, it misses the finer details,
such as phase reentrances and even the nature of the phase transition (discontinuous instead of continuous in some energy range)~\cite{nosnoneq}.

\begin{figure}[ptb]
\begin{center}
\scalebox{0.6}{{\includegraphics{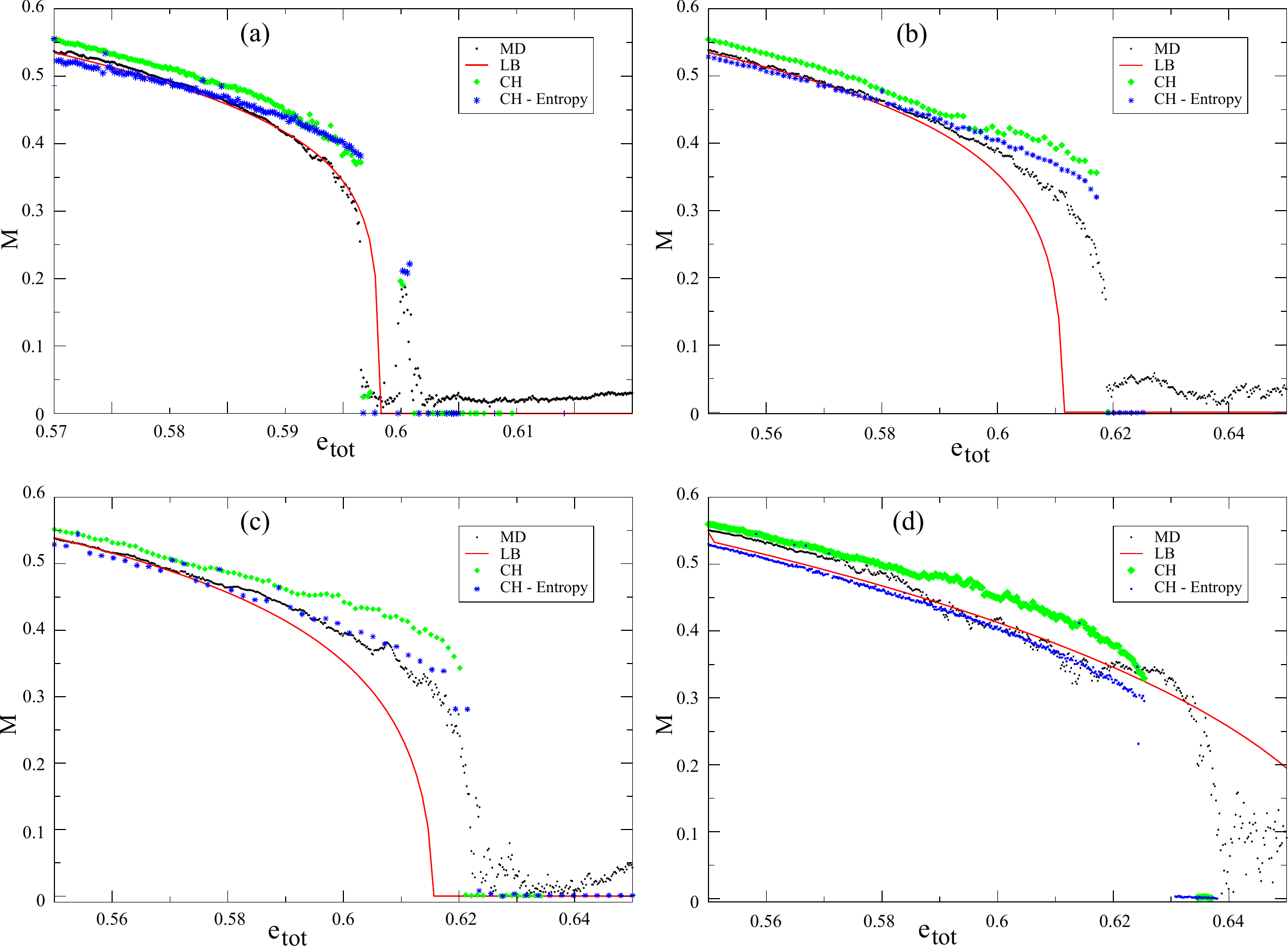}}}
\end{center}
\caption{(Color online) Final magnetizations for the HMF model as a function of total energy per particle $e_{\rm tot}$
from molecular dynamics (MD) simulations from $N=2\,000\,000$ particles and total simulation time $t_f=10\,000$,
from Lynden-Bell (LB), core-halo theory (CH) with halo energy $e_h$ obtained from the largest one-particle energy in the MD simulation,
and core-halo theory with $e_h$ with entropy maximization procedure (CH - Entropy).
Initial magnetizations are: a) $M_0=0.15$, b) $M_0=0.3$, c) $M_0=0.5$ and d) $M_0=0.8$.}
\label{fig4}
\end{figure}

\begin{figure}
\begin{center}
\scalebox{0.6}{{\includegraphics{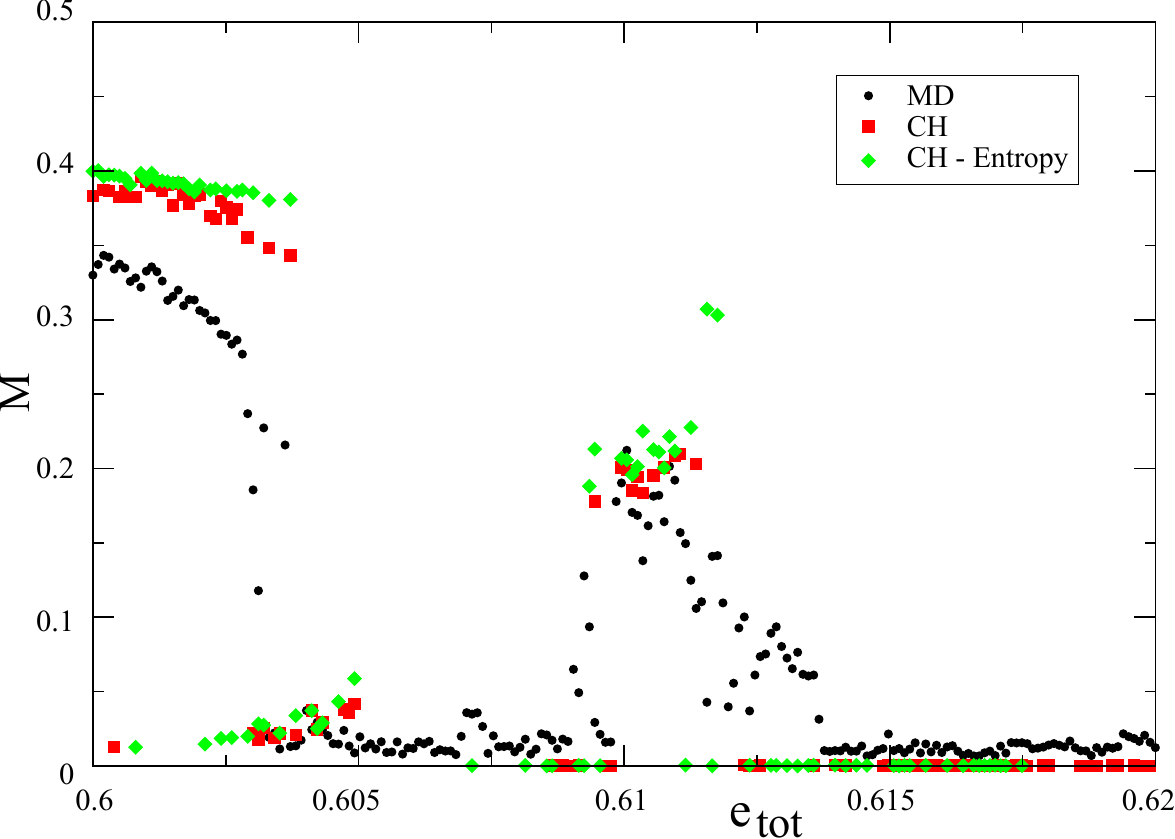}}}
\end{center}
\caption{(Color online) Zoom over Fig.~\ref{fig4}a with a close look at the phase reentrance details for initial magnetization $M_0=0.15$
and comparing MD simulations, core-halo (CH) and core-halo with entropy maximization (CH - Entropy)
as in Fig.~\ref{fig4}.}
\label{detailreen}
\end{figure}

The total energy of the system can also be written as
\begin{equation}
H=N\left[K+1-M_x^2-M_y^2\right],
\label{globalenerg}
\end{equation}
where $K$ is the kinetic energy per particle $K=(1/N)\sum_i p_i^2/2$
and $M_x$ and $M_y$ the magnetization components in the $x$ and $y$ directions, respectively:
\begin{equation}
M_x=\frac{1}{N}\sum_{i=1}^N\cos(\theta_i),
\label{defmagx}
\end{equation}
\begin{equation}
M_y=\frac{1}{N}\sum_{i=1}^N\sin(\theta_i).
\label{defmagy}
\end{equation}
The Fermi energy $e_F$ and the level of the halo $\chi$
are determined from the normalization condition:
\begin{equation}
\int \dd p\: \dd\theta f(p,\theta)=1,
\label{chnorm}
\end{equation}
and energy conservation
\begin{equation}
\int \dd p\: \dd\theta \left[\frac{p^2}{2}+\frac{1}{2}\,\overline{V}(\theta)\right]f(p,\theta)=e_{\rm tot},
\label{chenerg}
\end{equation}
where $e_{\rm tot}$ is the total energy divided by the particle number $N$.
The final magnetization is determined self-consistently as:
\begin{equation}
M=\int \dd p\: \dd\theta \cos(\theta)f(p,\theta),
\label{chmag}
\end{equation}
after setting the origin of the angles such that $M_y=0$ and $M_x=M\geq0$. The corresponding expression for the mean field potential is:
\begin{equation}
\overline{V}(\theta)=\int \dd p^\prime \dd\theta^\prime\left[1-\cos(\theta-\theta^\prime)\right]f(p^\prime,\theta^\prime)=1-M\cos(\theta).
\label{mfp0}
\end{equation}
Figure~\ref{fig4} shows, for a few different values of the initial magnetizations $M_0$, a comparison of MD simulations,
Lynden-Bell (LB) theory and the Core-Halo (CH) approach with the distribution in Eq.~(\ref{corehalodist}) and
$e_H$ determined from the highest one-particle energy from the same MD simulations.
Molecular dynamics simulations were performed using a graphics processing unit parallel implementation of a
fourth-order symplectic algorithm~\cite{yoshida,eu2}.
We note that due to the periodic boundary conditions the envelope equation is harder
to solve in the present case~\cite{levin4}. The CH approach predicts accurately the critical energy for the
ferromagnetic-paramagnetic transition in all cases but with quantitative discrepancies for the magnetization.
In the paramagnetic phase, the average magnetization from MD simulations does not vanish as a consequence of very-long lasting oscillations of
the spatial distribution around a QSS, the latter having a vanishing magnetization.
Is is important to note that such oscillations can last for a very long time or even forever~\cite{konishi}
and thus cannot be described by a static distribution function.
Nevertheless the vanishing average of each component of the magnetization are correctly predicted~\cite{levin4,nosnoneq}.
In Fig.~\ref{fig4}a a phase reentrance is clearly visible and its position is accurately predicted by the CH approach.
A closer look at the phase reentrance region is given in Fig.~\ref{detailreen}.
Some deviations from the numerical values of $M$ are due to the fact that the ansatz
in Eq.~(\ref{corehalodist}) is too simple and cannot grasp all the details of a more complex distribution function
(see Fig.~\ref{fig5} below).

\section{Variational method: Entropy maximization}
\label{varmethod}

The very long relaxation time to thermodynamic equilibrium of a system with long-range interactions, which diverges with $N$~\cite{physrep,dynamics,scaling},
is a consequence of the existence of the infinitely many Casimir constants of motion of the Vlasov equation~(\ref{vlasoveq}).
Those are usually incompatible with the values of the Boltzmann equilibrium distribution ${\cal C}\exp[-\beta e({\bf r},{\bf p})]$,
with ${\cal C}$ a normalization constant and $e({\bf r},{\bf p})$ given in Eq.~(\ref{mfenergy}). Consequently
the system can never attain equilibrium in the limit $N\rightarrow\infty$, and eventually settles in a
stationary state of the Vlasov equation (or oscillates around it).
For finite $N$, collisional effects (graininess) become important and the system is in a QSS which
relaxes very slowly to equilibrium~\cite{dynamics}.
In Ref.~\cite{nosprl} it was shown that the stability conditions for a homogeneous QSS of the HMF
model as obtained in Ref.~\cite{nv1} is equivalent to maximizing the Gibbs entropy subject to the constraints of energy conservation,
normalization and all the analytic Casimirs (i.e.\ with an analytic function $s$ in Eq.~(\ref{casimirs})).
On the other hand, it is a well known property of the Vlasov equation that the dynamical evolution leads to the
formation of filaments in a scale that gets smaller with time, and which
leads to difficulties in its numerical integration due to the finite precision of a numerical grid~\cite{eu1}.
For a description using individual particle dynamics, the finite computer precision also introduces a loss of information of the details of the filamentation.
In both cases this amounts to a coarse-graining no matter the grid resolution, and results in an increase of the Gibbs entropy corresponding to
$s=-f\ln f$ in Eq.~(\ref{casimirs}) but computed using a coarse-grained distribution $f_{cg}$:
\begin{equation}
S_G=-\int\dd{\bf p}\:\dd{\bf r}\: f_{cg}({\bf p},{\bf r};t)\ln f_{cg}({\bf p},{\bf r};t),
\label{gibbsent}
\end{equation}
where for simplicity and from now on we set the Boltzmann constant to unity.
\emph{From now on, we will only deal with the coarse grained one-particle distribution function and all Casimirs will be considered relative to 
this same distribution}.

Figure~{\ref{figentropy} shows $S_G$ for a few different grid resolutions and a
waterbag initial condition. In our approach we consider
indirectly the effective values of the Casimirs as given by the coarse-grained description.
The size of the coarse-graining does not affect the resulting distribution function, provided it is sufficiently small,
as it becomes evident from
Fig.~\ref{figcasimirs} showing the Casimirs $C^{(k)}$ for a few values of $k$
and the same initial condition. A similar behavior is observed for other choices of $s$.
We observe that no matter the grid resolution, the final values of $S_G$ and $C^{(k)}$ are the same, up to small numerical errors.
It is beyond the scope of the present paper to discuss the effects of a coarse-graining of the dynamics of the system which is discussed in greater detail in
Refs.~\cite{firpo1,doveil}. The discussion above was intended to show that for the present purpose the details of the coarse-graining are not relevant
in the determination of the statistical state of the system after a violent relaxation.

The values reached by the coarse-grained Casimirs, after the system has settled in a QSS, uniquely define
the one-particle distribution function if we suppose that the latter depends only on the energy,
as it is true for one-dimensional systems from Jeans theorem, provided some assumptions are met~\cite{binney}.
To show that the Casimirs determine the distribution it is sufficient to consider Casimirs of the form:
\begin{equation}
C^{(k)}=\int \dd{\bf p}\:\dd{\bf r} \left[f({\bf p},{\bf r})\right]^k,
\label{defcask}
\end{equation}
with $k$ a positive integer. Let us consider as a first approximation a distribution function with $L$ discrete values
$f_i$, $i=1,\ldots,L$ such that $f({\bf r},{\bf p})=f(e({\bf r},{\bf p}))=f_i$
in the one-particle phase space region $\omega_i$ defined by $e_{i-1}<e({\bf r},{\bf p})\leq e_i$.
By taking the limit of an infinite number of levels for $f$ we recover a continuous function.
In this way the Casimirs in Eq.~(\ref{defcask}) are rewritten as
\begin{equation}
C^{(k)}=\sum_{i=1}^L f_i^kS_i,
\label{defcask2}
\end{equation}
where $S_i$ is the volume of the region $\omega_i$. It is important to note that the value of $C^{(k)}$ is obtained using the coarse-grained distribution function.
The values of $f_i$ and $S_i$ are then specified by fixing the values of a sufficient number of Casimirs.

The level curves of constant one-particle mean field energy $e({\bf r},{\bf p})$ are  obtained
from the values of $S_i$. This is equivalent to:
\begin{equation}
e({\bf r},{\bf p})=\frac{p^2}{2}+\overline{V}({\bf r})=e_i,\hspace{5mm}i=1,\ldots,k,
\label{prf0}
\end{equation}
with
\begin{equation}
\overline{V}({\bf r})=\sum_{i=1}^kf_i\int_{\omega_i} \dd{\bf p}^\prime\:\dd{\bf r}^\prime\:V({\bf r}-{\bf r}^\prime)=\sum_{i=1}^kf_i V_i({\bf r}),
\label{prf1}
\end{equation}
and
\begin{equation}
V_i({\bf r})\equiv\int_{\omega_i} \dd{\bf p}^\prime\:\dd{\bf r}^\prime\:V({\bf r}-{\bf r}^\prime),\hspace{5mm}
S_i=\int_{\omega_i} \dd{\bf p}^\prime\:\dd{\bf r}^\prime.
\label{prf2}
\end{equation}
Each value of $i$ in Eq.~(\ref{prf0}) then defines the contour curves (or surfaces) of each region $\omega_i$.
This last equation defines self-consistently the boundary of each region $\omega_i$ from the set of points $({\bf r},{\bf p})$ satisfying it.

Let us illustrate how this procedure with the HMF model with a two-level distribution function ($L=2$) as in the core-halo ansatz in
Eq.~(\ref{mfp0}), such that $e_1=e_F$ and $e_2=e_H$. The mean-field potential is given by Eq.~(\ref{mfp0}), and from Eq.~(\ref{prf0}) we have
\begin{equation}
\frac{p^2}{2}=1-M\cos(\theta)=e_i,
\label{eqn1}
\end{equation}
with solution
\begin{equation}
p=\pm\sqrt{2}\sqrt{e_i-1+M\cos(\theta)},
\label{eqn2}
\end{equation}
which defines the frontier curves of $\omega_1$ and $\omega_2$. Note that depending on the one-particle energy these curves
can be composed by two disjoint curves (in the case of the HMF model), but the discussion in the previous paragraph still holds in this case.
Indeed, it is straightforward to show that no two different such curves can have the same set of Casimirs of the form discussed above.
The magnetization $M$ is obtained self-consistently from Eq.~(\ref{chmag}).

The dynamics drives the system through a violent relaxation which then settles into a QSS. If the values of the Casimirs are known, then
the distribution function can be determined. The values of the (coarse-grained) Casimirs can be directly determined from the core-halo ansatz
in Eq.~(\ref{corehalodist})
as a function of the parameters $\eta$, $\chi$, $e_F$ and $e_H$, which can be determined from the value of $f=\eta$ in the initial condition,
energy conservation and normalization of $f$. There still remains to determine the halo energy parameter $e_H$.
It is much natural to expect that the system, if no dynamical constraints forbid so, evolves into the most probable state
In statistical inference, the most probable state is determined by maximizing the Gibbs entropy in Eq.~(\ref{gibbsent})
modulo any constraints. Here we only have to consider the constraints of total energy and normalization of $f$.
The values of the Casimirs as a function of the free parameters are already given by the core-halo ansatz.
Note that the Gibbs entropy is the only additive form with no statistical bias, up to a multiplicative and an additive constants,
meaning that in the absence of any constraint, all states are equally probable.
Any other form of the entropy than $S_G$ leads to a bias (see for instance the discussion in Ref.~\cite{nosent}).
At this point it is important to note that the most probable state is unique, up to some degeneracies related to conserved quantities,
in the same way as the thermodynamic equilibrium
is the unique state obtained as the most probable state given the total energy of the system.

In order to corroborate this statement, the Gibbs entropy
as a function of the halo energy $e_H$ for $M_0=0.15$ and $e=0.61$ is shown in Fig.~\ref{ent1}.
The values of $e_H$ obtained as the highest particle energy and the maximum of $S_G$ differ only in the second decimal digit.
The non-equilibrium phase diagram for different initial magnetizations obtained from the variational approach with the ansatz
in Eq.~(\ref{corehalodist}) are shown in Fig.~\ref{fig4}. We observe that they are very close and even slightly better than those
obtained from the original core-halo method.

\begin{figure}[ptb]
\begin{center}
\scalebox{0.3}{{\includegraphics{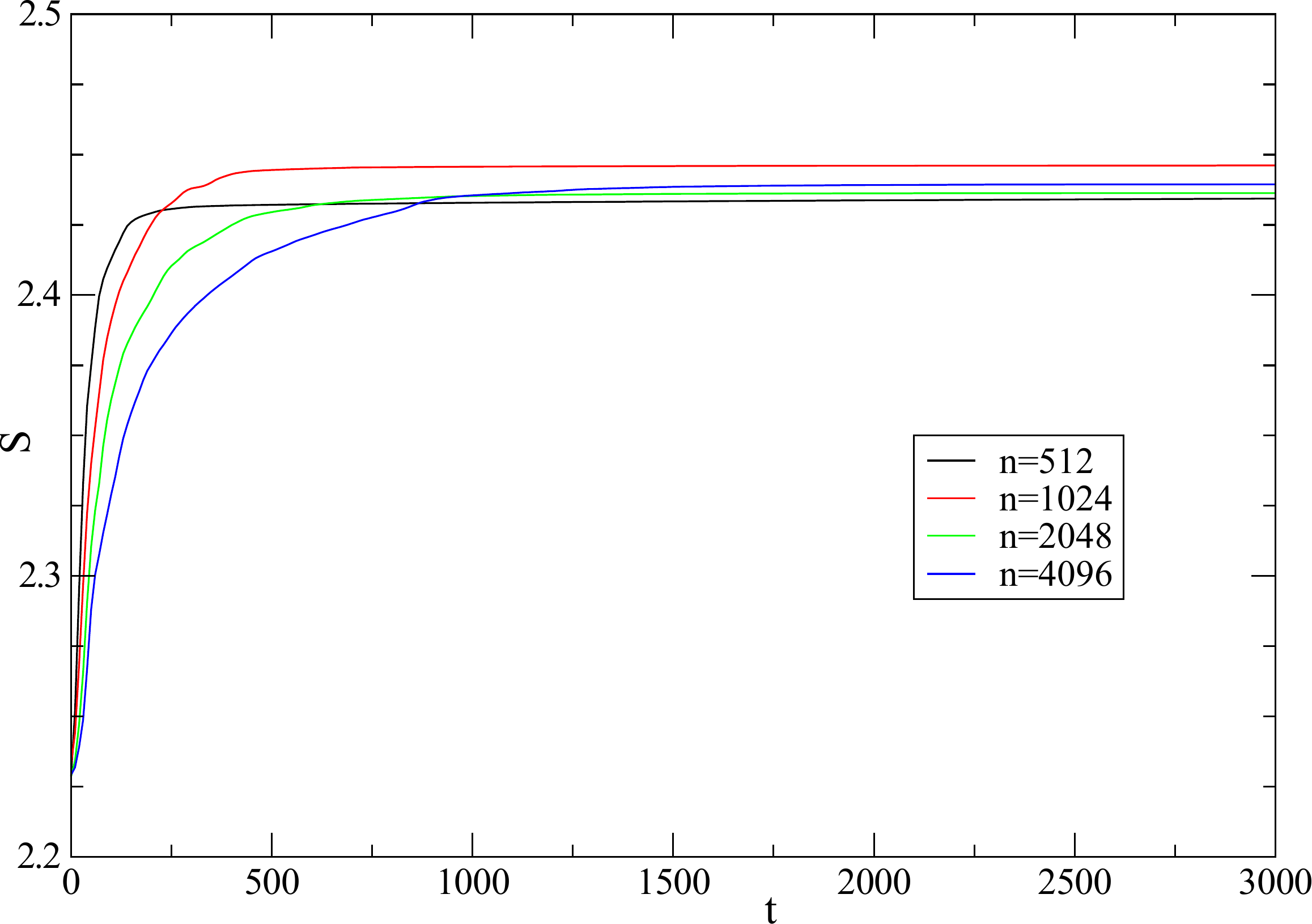}}}
\end{center}
\caption{Entropy from the solution of the Vlasov equation as a function of time for $M_0=0.15$ and $e_{\rm tot}=0.61$ with different numerical grid resolutions
$n_p\times n_\theta=n$, where $n_p$ and $n_\theta$ are the number of points in the momentum and position directions respectively. The momentum
varies in the interval $[-2.56,2.56]$ and position from $0$ to $2\pi$.}
\label{figentropy}
\end{figure}

\begin{figure}[ptb]
\begin{center}
\scalebox{0.3}{{\includegraphics{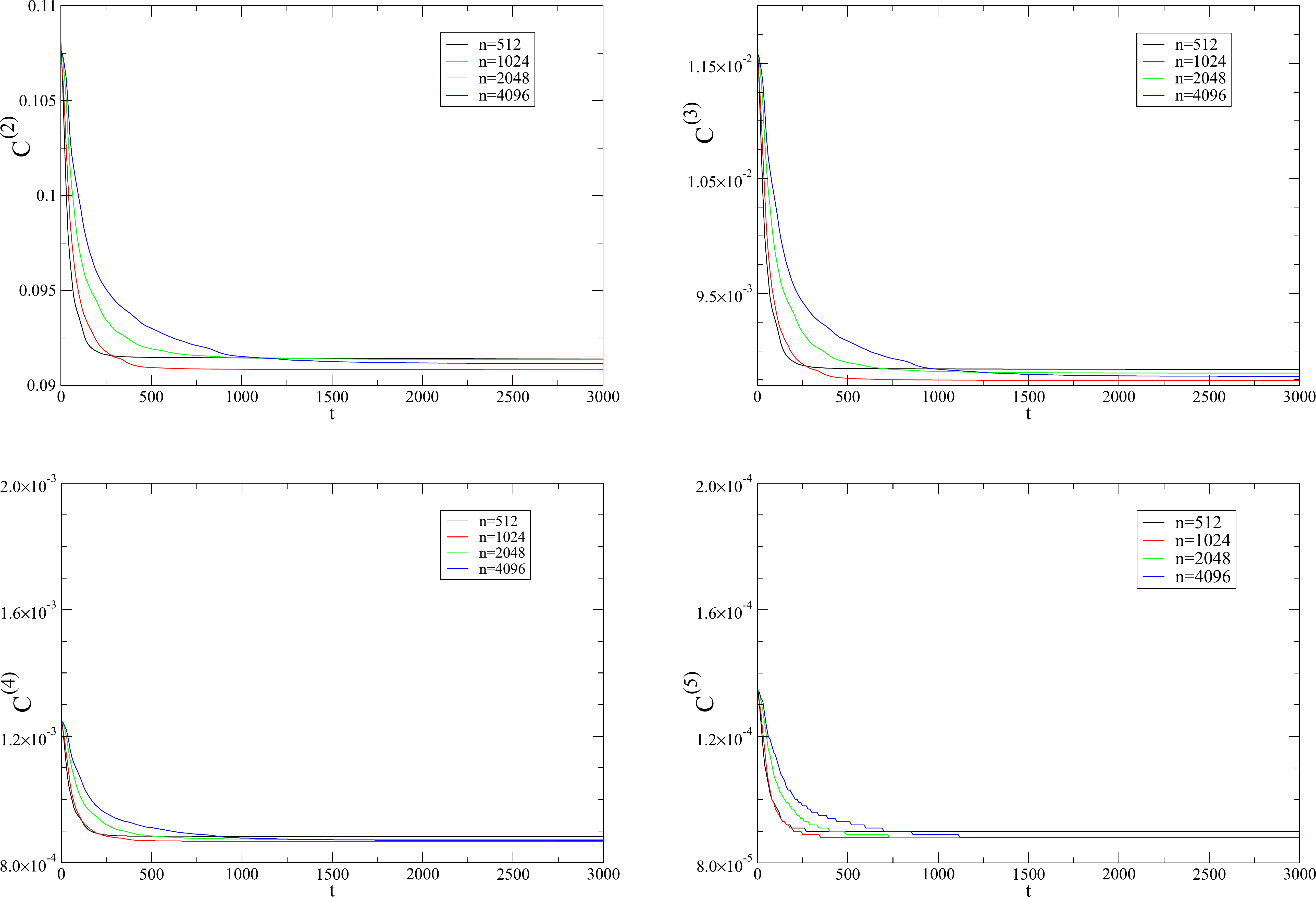}}}
\end{center}
\caption{Casimirs $C^{(k)}$ for $k=2,3,4,5$ as defined in Eq.~(\ref{defcask}) as a function of time from the solution of the Vlasov
equation for the same initial condition and grid resolutions as in Fig.~\ref{figentropy}.}
\label{figcasimirs}
\end{figure}

\begin{figure}[ptb]
\begin{center}
\scalebox{0.6}{{\includegraphics{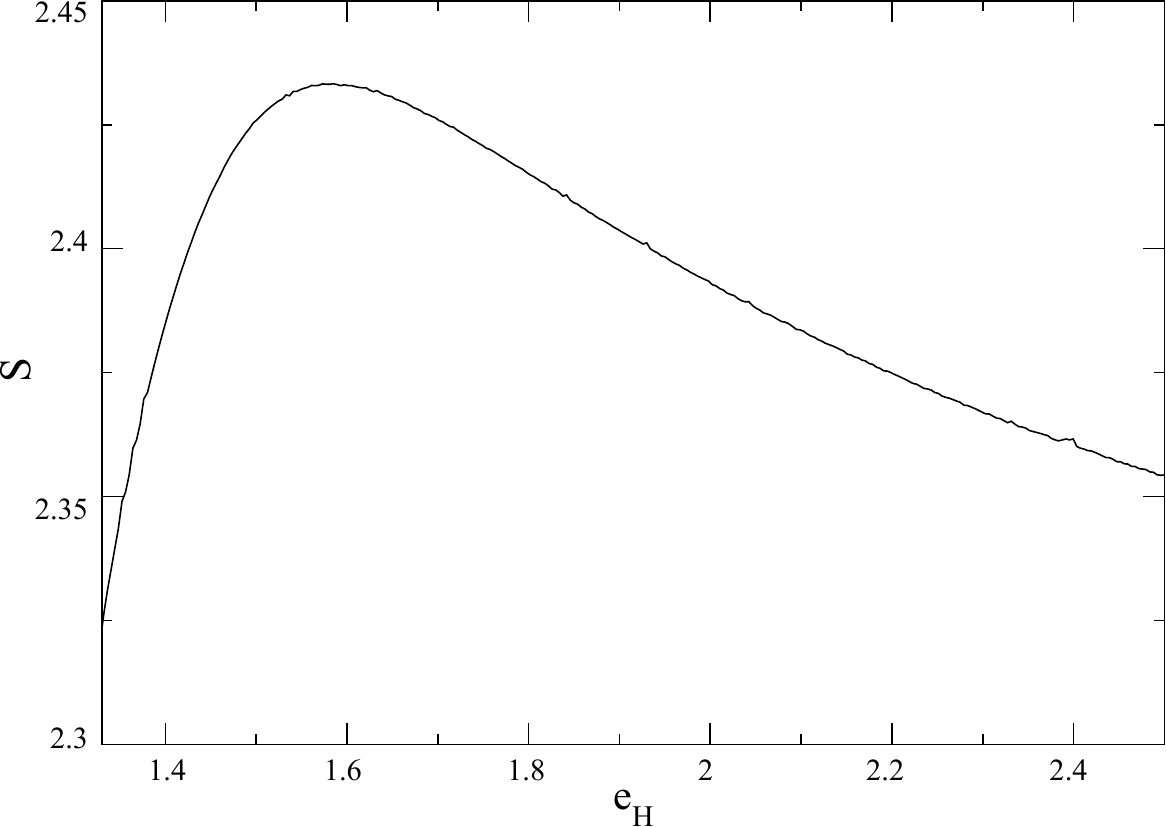}}}
\end{center}
\caption{Gibbs entropy as a function of $e_H$ for $M_0=0.15$ and $e=0.61$. The maximum of $S_B$ is at $e_h=1.573$,
which is very close to the value $e_H=1.535$ obtained from the particle with maximum energy in an MD simulation.}
\label{ent1}
\end{figure}

A variational method opens the way to use a different ansatz as a trial function to determine the extremum of the considered functional.
Here the choice of ansatz is dictated from simulation results which consistently show a core-halo structure~\cite{levipprep}.
The dependence of the distribution function $f(\theta,p)$ on the mean-field one-particle energy $e(\theta,p)$
can be obtained numerically, as shown in Fig.~\ref{fig5} for some values of the total energy per particle $e_{\rm tot}$
and with initial magnetization $M_0=0.15$. The core-halo distribution function as obtained from both the original
and the variational method (i.e.\ with $e_H$ determined from entropy maximization) are also shown. Although the separation in a core and a halo becomes less
evident as the energy augments, both yield very similar results. From Fig.~\ref{fig5}
it is quite natural to try a different ansatz given by:
\begin{equation}
f({\bf r},{\bf p},t_v)=\eta\,\Theta(e_F-e({\bf r},{\bf p}))+\chi\Theta(e({\bf r},{\bf p})-e_F)\Theta(e_H-e({\bf r},{\bf p}))
\left[1-\frac{e({\bf r},{\bf p})-e_F}{e_H-e_F}\right],
\label{corehalomod}
\end{equation}
which is essentially a core with constant $f$-value $\eta$, also given by the value of the initial waterbag distribution,
but with a linearly decreasing halo starting at the $f$-value $\chi$.
The resulting values of magnetization as a function of $e$ after the violent relaxation are shown in Fig.~\ref{fig8}
with a little improvement for some energy intervals.
The ansatz in Eq.~(\ref{corehalomod}) is closer to the dependence of the distribution functions in the one-particle energy $e$
as obtained from MD simulations,
at least for the cases considered here, as can be seen in Fig.~\ref{fig5}. The velocity and position distribution functions for the different approaches
considered here are given in Fig.~\ref{vthdist}, with little differences from one another.
The Fermi and halo energies for the case $M_0=0.15$ are shown in Fig.~\ref{fig6}, with substantial differences. We note that in our results
the value of $e_F$ and consequently also that of $e_H$ are bigger than the average total energy per particle, indicating that the majority of particle are
concentrated on the core of the distribution.
Our results also show that relevant physical observables have little sensitivity in the details of the halo part of the distribution.

The (non-equilibrium) entropy for the usual core-halo distribution in Eq.~(\ref{corehalodist}) is shown in Fig.~\ref{fig10} as a function of energy $e$ and $e_h$ for $M_0=0.15$.
The discontinuous phase transition is related to a discontinuity in the entropy at the critical energy and, quite interestingly, the entropy decreases
in the phase transition, as energy increases, which would be impossible for an equilibrium phase transition.
This means that at the onset of the parametric resonance that triggers the halo formation and the phase transition,
the region in the $(\theta,p)$ single particle phase space of states accessible to the system and compatible with all constraints (energy and Casimir invariants)
in fact shrinks at the phase transition, while the energy increases. By comparing Fig.~\ref{fig10} and Fig.~\ref{detailreen} we note the complicate phase reentrance
structure observed in the latter is also associated to discontinuities in the non-equilibrium Gibbs entropy resulting from the core-halo ansatz.

Although all expressions used here are typically analytic in the free parameters $\chi$, $e_F$, $e_H$ and $M$, the resulting non-equilibrium entropy has discontinuities at 
some values of the single particle energy $e$. This is due to the fact that
$\chi$, $e_H$ and the magnetization $M$ are obtained from the solution of highly nonlinear equations. The solutions of the latter can present discontinuities when
a given parameter changes, as the total energy $e$ or the initial magnetiztion $M_0$.

The interplay between Gibbs statistics and dynamical regimes has already been studied previously by many authors in both short- and long-range interacting
systems~\cite{escande,baldovin,tamarit,yang}. Here we have shown evidence of a similar interplay 
between a dynamic property (parametric resonance) and a quasi-stationary non-equilibrium statistical distribution.

\begin{figure}[ptb]
\begin{center}
\scalebox{0.6}{{\includegraphics{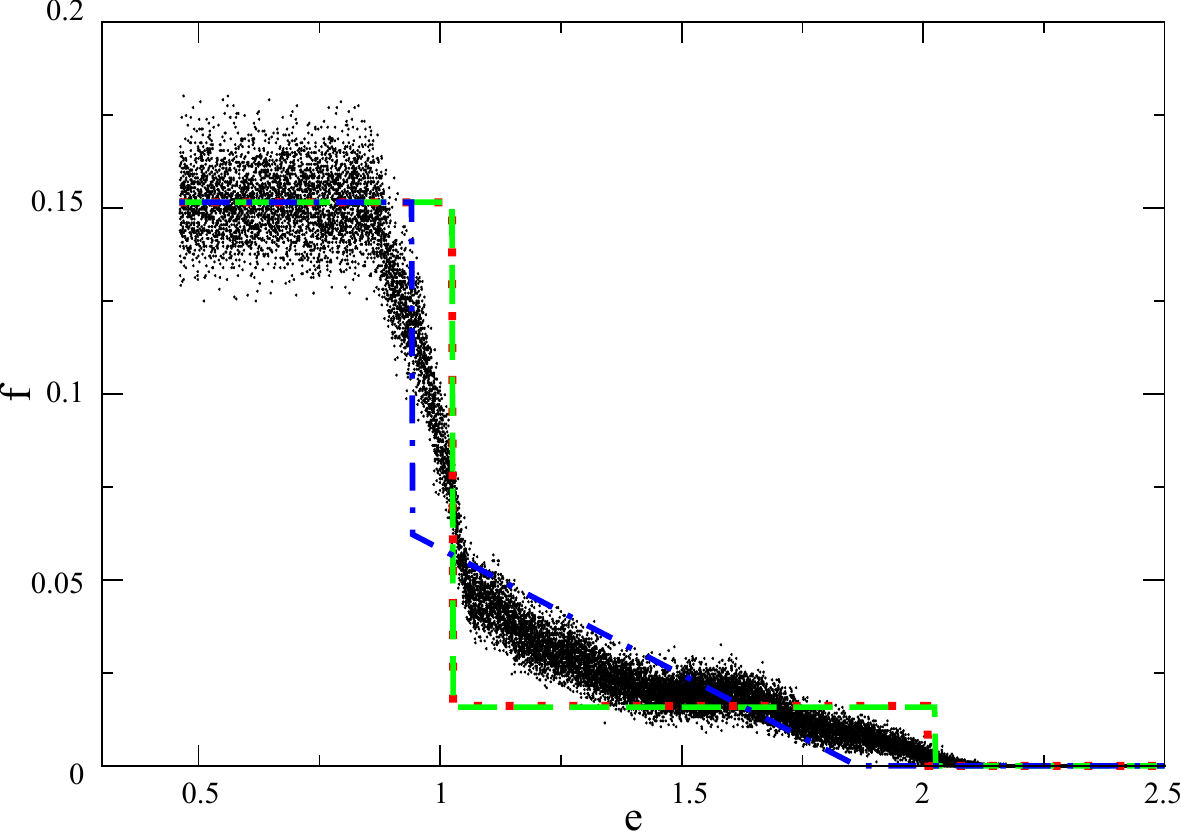}}}
\scalebox{0.6}{{\includegraphics{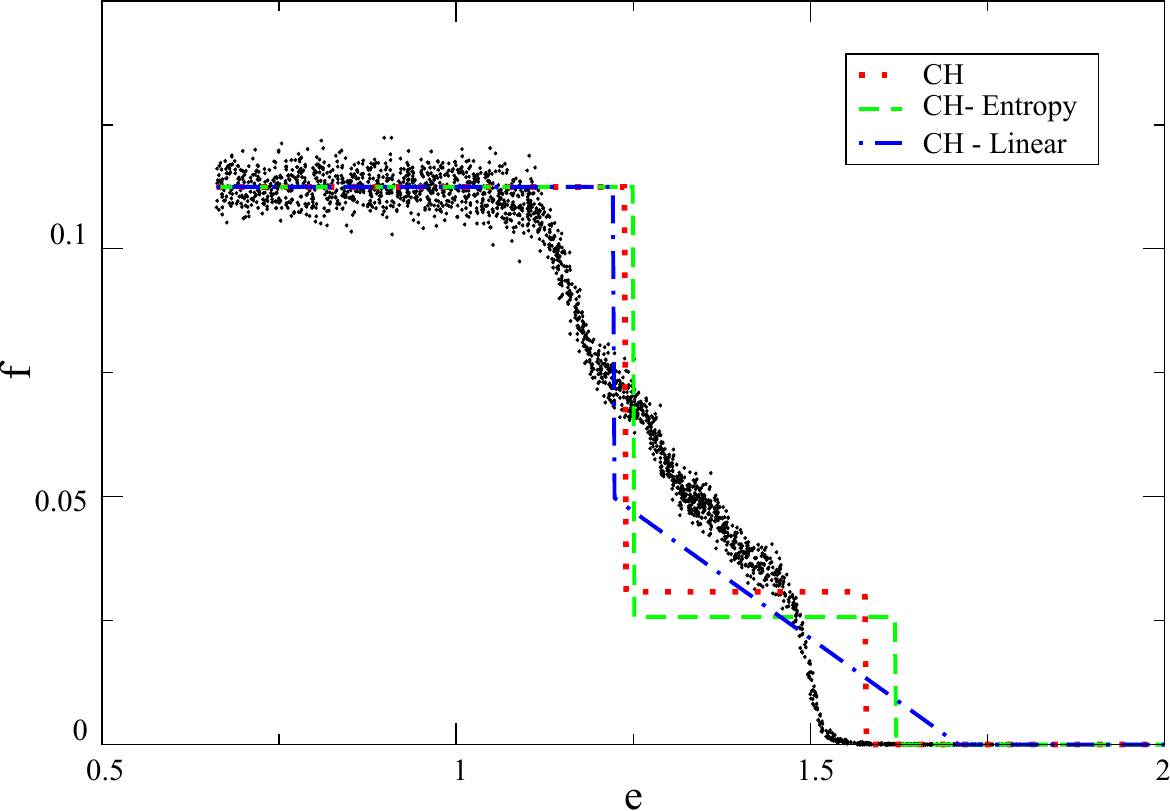}}}
\scalebox{0.6}{{\includegraphics{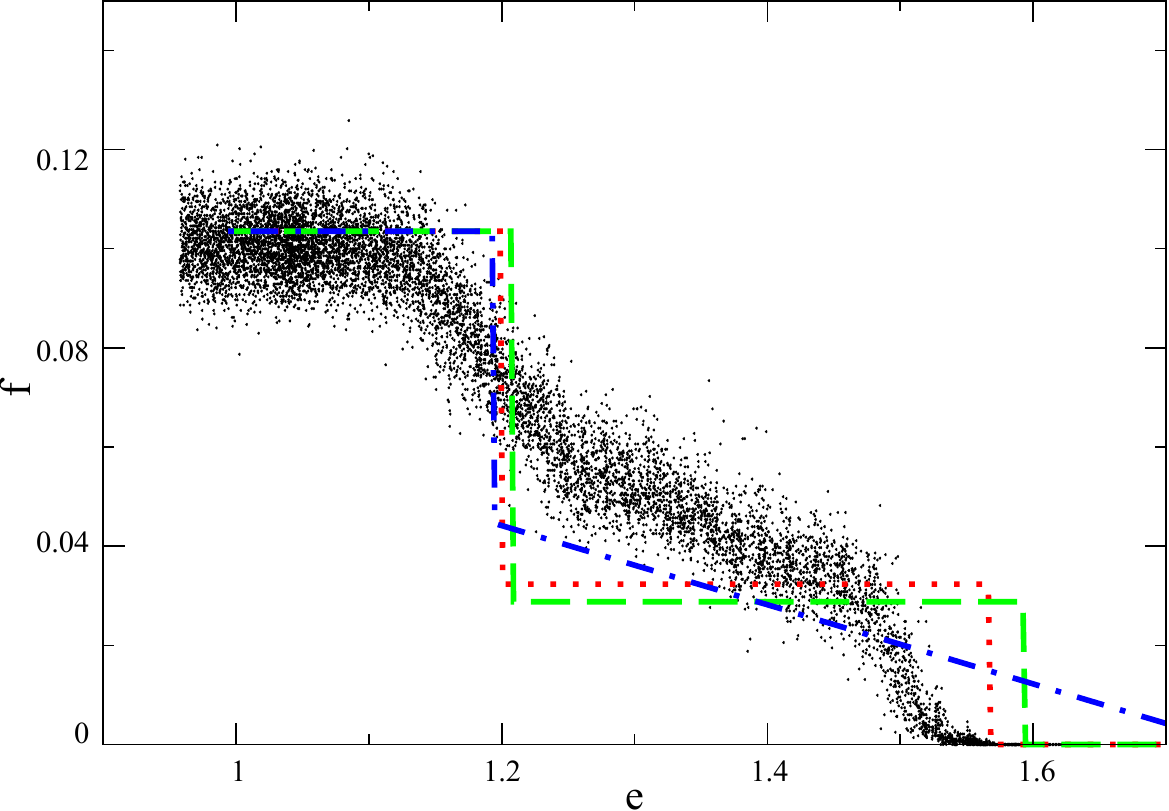}}}
\end{center}
\caption{Distribution functions $f(e(p,\theta))$ for the HMF model
as a function of single particle energy for initial magnetization $M_0=0.15$ and total energies per particle
a) $e_{\rm tot}=0.55$, b) $e_{\rm tot}=0.6$, c) $e_{\rm tot}=0.62$.
Total simulation time is $t_f=100\:000$ in order to allow a more complete thermalization and $N=2\,000\,000$ (initial condition as in Fig.~\ref{fig4}),
except for (c) with $N=20\,000\,000$ and $t_f=10\,000$.
The result obtained from the original core-halo approach (CH), from the present variational method (CH-Entropy) and from the modified ansatz
in Eq.~(\ref{corehalomod}) (CH-Linear) are also shown.}
\label{fig5}
\end{figure}

\begin{figure}[ptb]
\begin{center}
\scalebox{0.6}{{\includegraphics{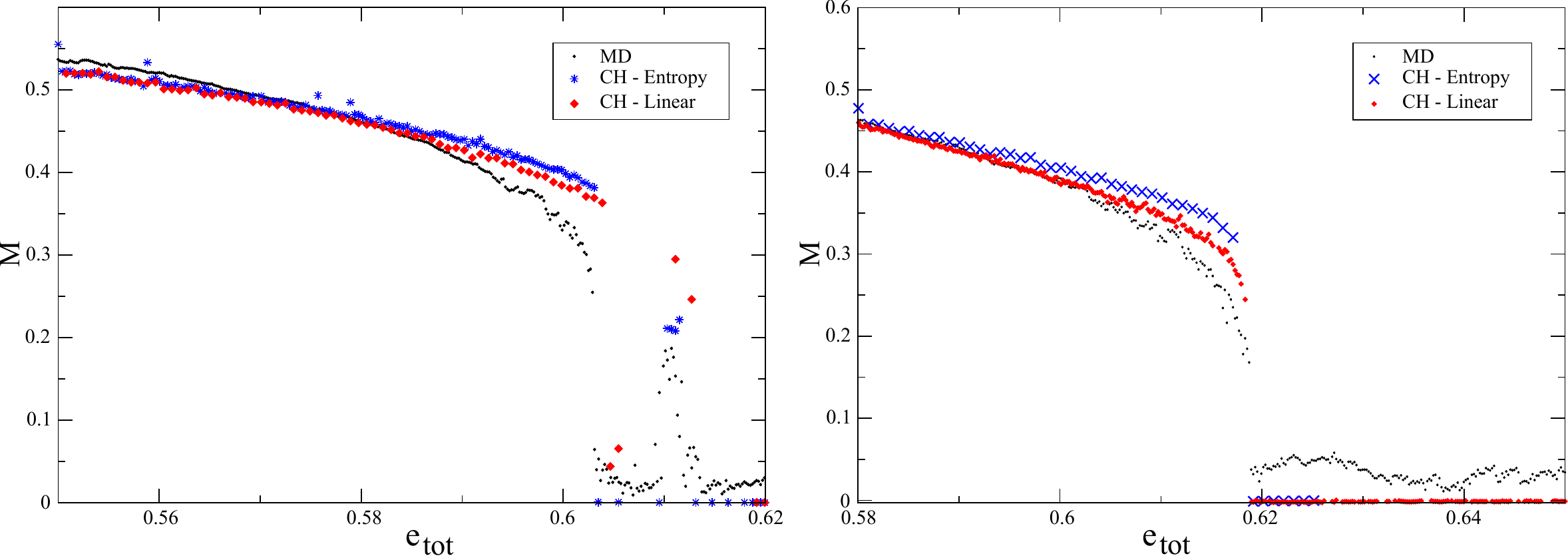}}}
\end{center}
\caption{(Color online) Final magnetization after the violent relaxation as a function of energy for the original core-halo (CH-Entropy)
and the modified ansatz in Eq.~(\ref{corehalomod}) (CH - Linear), where in both cases $e_H$ was determined from the entropy maximum.
Initial magnetizations are: a) $M_0=0.15$ b) $M_0=0.3$. The values obtained from the distribution function in Eq.~(\ref{corehalodist}) are
also plotted for comparison.
Both cases were obtained maximizing the entropy.}
\label{fig8}
\end{figure}

\begin{figure}[ptb]
\begin{center}
\scalebox{0.6}{{\includegraphics{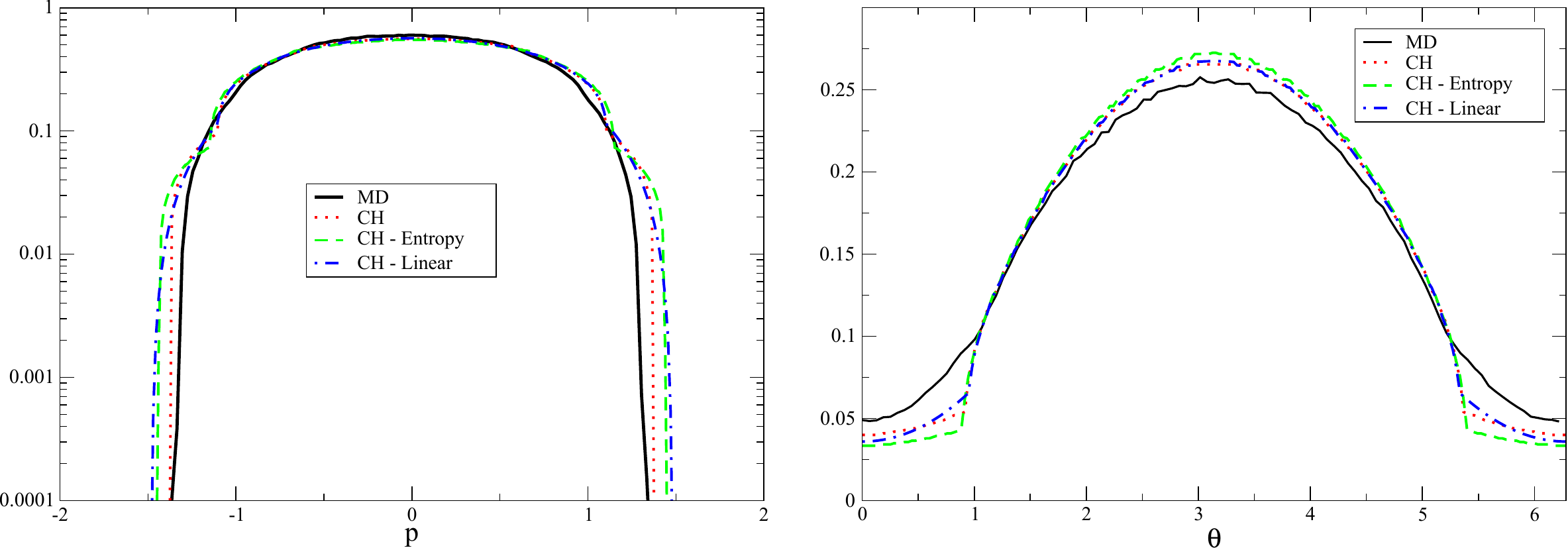}}}
\end{center}
\caption{(Color online) Left panel: Mono-Log plot of the velocity distribution function from molecular dynamics (MD), the original core-Halo distribution (CH),
the core-halo distribution from entropy maximization (CH-Entropy), and the alternative core-halo distribution with the linearly decreasing halo in Eq.~(\ref{corehalomod})
and maximizing the entropy (CH-Linear) for $M_0=0.15$ and $e=0.6$. Right panel: The spatial distribution function for the same cases.}
\label{vthdist}
\end{figure}

\begin{figure}[ptb]
\begin{center}
\scalebox{0.6}{{\includegraphics{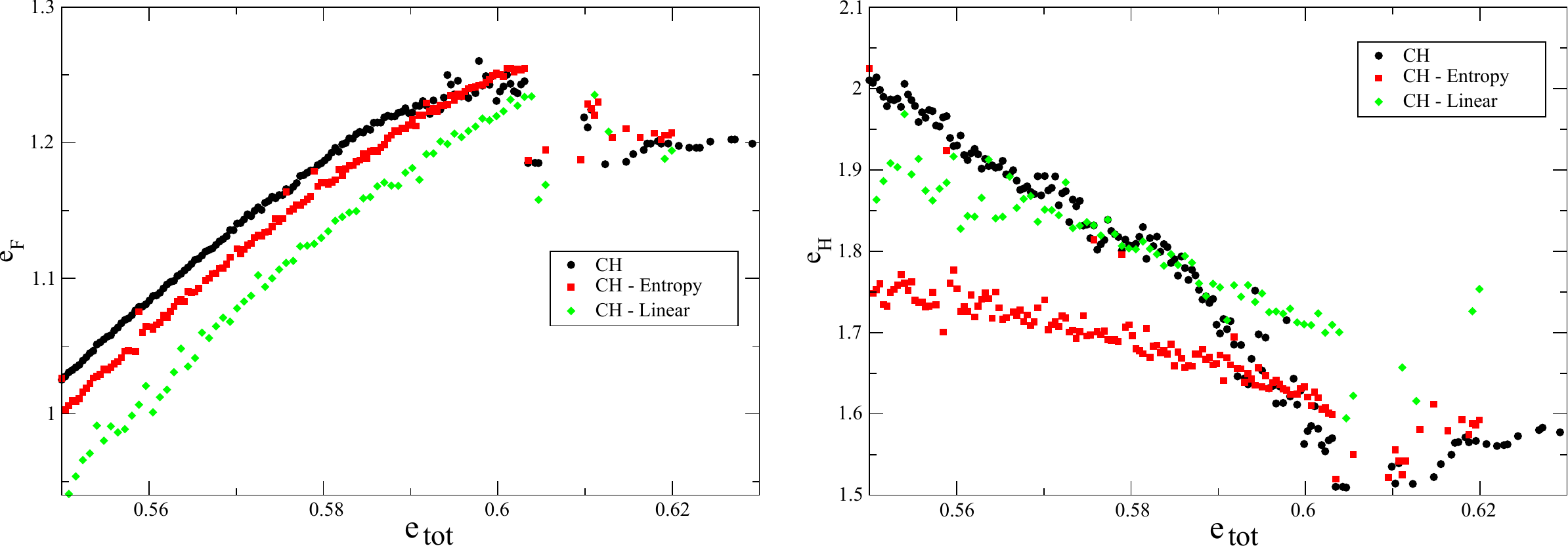}}}
\end{center}
\caption{(Color online) Fermi energy $e_F$ and halo energy $e_H$ for $M_0=0.15$ as a function of total energy per particle
for the Core-Halo approach (CH) with $e_H$ obtained from MD
simulations, with $e_H$ obtained form entropy maximization (CH - Entropy),
and from the modified Core-Halo function in Eq.~(\ref{corehalomod}) and entropy maximization
(CH-Linear).}
\label{fig6}
\end{figure}

\begin{figure}[ptb]
\begin{center}
\scalebox{0.3}{{\includegraphics{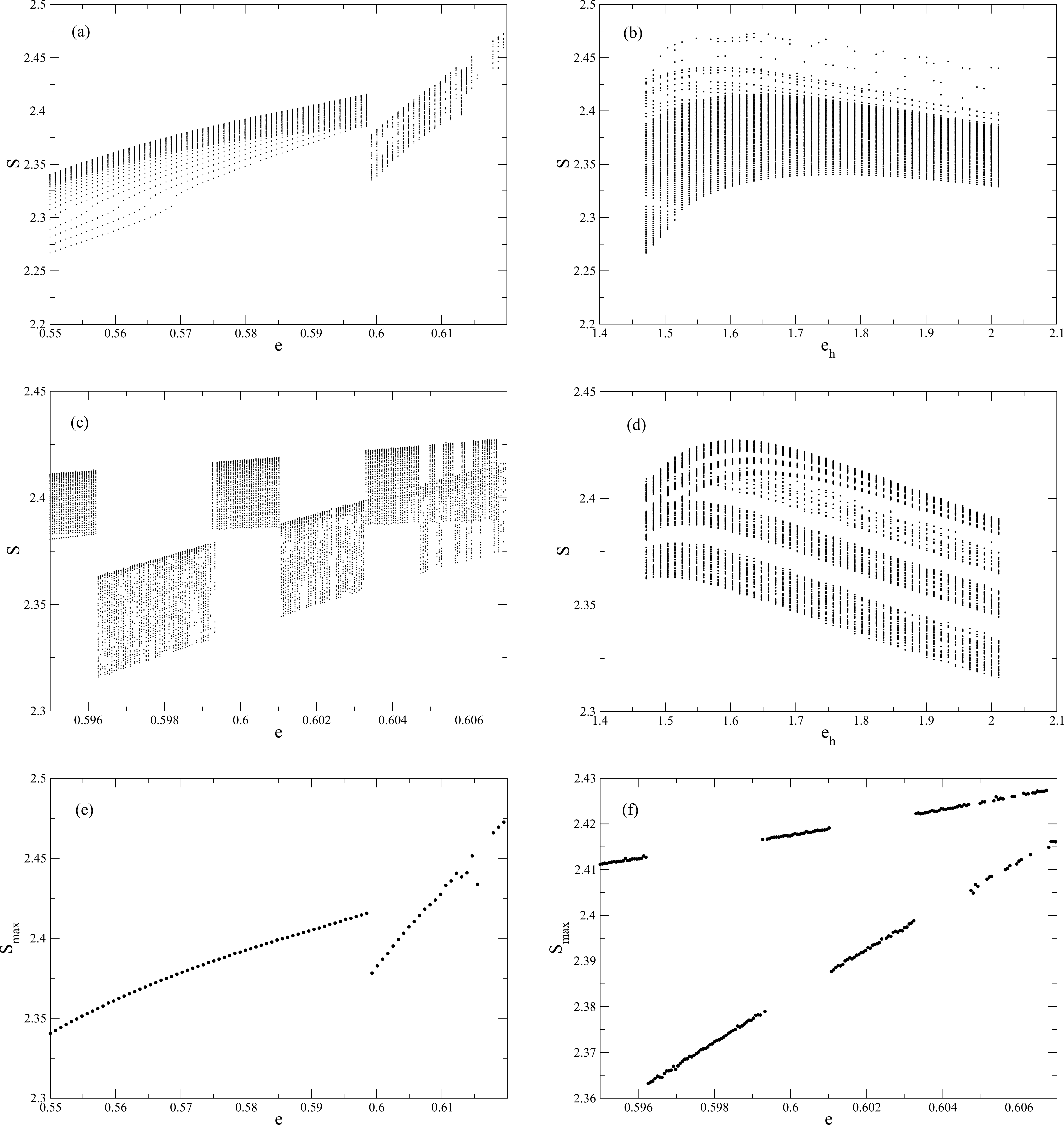}}}
\end{center}
\caption{a) Gibbs entropy in Eq.~(\ref{gibbsent}) for the core-halo distribution in Eq.~(\ref{corehalodist}) as a function of total energy per particle $e$
for different values of the halo energy $e_h$ in the interval $1.46<e_h<2.0$ and initial magnetization $M_0=0.15$.
b) Entropy as a function of the core-halo energy $e_h$ for the interval of energy values $e$ in (a).
c) Same as (a) but zooming around the energy value of the phase transition. d) Same as (b) but with an interval for $e$ around the phase transition.
e) Maximum Gibbs entropy $S_{max}$ as a function of $e$. f) Same as (e) but zooming on the energy value of the phase transition. Note that phase reentrances
also correspond to discontinuities of the entropy.}
\label{fig10}
\end{figure}

\section{Concluding Remarks}
\label{conclusions}

We have shown that the core-halo approach of Levin and collaborators can be recast as a variational principle when the Gibbs entropy maximization principle
is considered alongside the energy and mass conservation (normalization of $f$), with the final magnetization obtained self-consistently.
The information on the Casimir values is in fact embedded in the ansatz for the core-halo distribution and maximizing the entropy
is then equivalent to stating that the system evolves towards the most probable state given the energy, initial magnetization and
the information on the final values of the coarse-grained Casimirs in the core-halo ansatz. The results obtained using the present variational approach
are compatible with those obtained previously for the HMF model, and slightly better for some values of energy and initial magnetization.
We also obtained, for the first time using a theoretical approach, the phase-reentrances of the HMF model, and showed that it also results from discontinuities
of the entropy that result straightforwardly from the information contained in the core-halo distribution.
It is quite interesting that in the non-equilibrium case the phase transition is not always due to a discontinuity
on a derivative of the entropy, but of the entropy itself.

We also shown that other forms for the distribution function as a function of the one-particle energy can be used with similar results,
provided the number of parameters is the same. This is in fact a common advantage of variational methods, allowing the use of different trial functions in the functional
to maximize. Our approach greatly simplifies the application of the core-halo approach as no envelope equation
is required to determine the halo energy. Although the present approach works well with HMF systems and is robust if one admits that the system
must evolve towards the most probable states given a set of constraints, it must still be applied to other long-range interacting systems in order to assess its general
validity, which is the subject of ongoing work.

Another important result of the present paper concerns the link between an out-of-equilibrium dynamical phenomenon, the parametric resonance
causing the non-equilibrium phase transition, and the entropy as a statistical property of a (non-equilibrium) stationary state. The dynamics
leads the system into its final state (in the Vlasov limit $N\rightarrow\infty$) and, in some still undetermined way, must have a signature in the entropy maximum
corresponding to this state. Although the core-halo approach is not a complete theory for the determination of the outcome of the violent relaxation evolution,
the present work shows the relevance of statistical concepts such as entropy maximization subject to dynamical constraints in the form of Casimir and
energy conservation.

\section{Acknowledgments}

The author acknowledges partial financial support by CAPES and CNPq (Brazilian Government agency).

\end{document}